\PassOptionsToPackage{dvipsnames}{xcolor}
\documentclass[10pt, conference]{IEEEtran}
\IEEEoverridecommandlockouts
\usepackage[hidelinks]{hyperref}
\usepackage{verbatim}
\usepackage{amsmath,amssymb,amsfonts}
\usepackage{textcomp}
\def\BibTeX{{\rm B\kern-.05em{\sc i\kern-.025em b}\kern-.08em
    T\kern-.1667em\lower.7ex\hbox{E}\kern-.125emX}}

\newcommand{\ignore}[1]{}

\usepackage{ulem}
\usepackage[switch]{lineno}

\usepackage[T1]{fontenc}
\usepackage{listings}
\usepackage{xcolor}
\definecolor{verylightgray}{rgb}{.97,.97,.97}
\lstdefinelanguage{FunC}{
  keywords=[1]{if, ifnot, else, elseif, elseifnot, while, do, until, repeat, return, impure, method_id, forall, asm, inline, inline_ref}, 
  keywordstyle=[1]\color{blue}\bfseries,
  keywords=[2]{var, int, slice, tuple, cell, builder, cont, global, const, _},  
  keywordstyle=[2]\color{teal}\bfseries,
  keywords=[3]{cur_lt, block_lt, now, my_address, get_balance, config_param, get_data, set_data, random, rand, get_seed, set_seed, randomize, randomize_lt, load_msg_addr, parse_addr, parse_std_addr, parse_var_addr},  
  keywordstyle=[3]\color{violet}\bfseries,
  keywords=[4]{begin_cell, end_cell, begin_parse, end_parse, load_ref, preload_ref, load_int, load_uint, preload_int, preload_uint, load_bits, preload_bits, load_coins, skip_bits, first_bits, skip_last_bits, slice_last, load_dict, preload_dict, skip_dict, store_ref, store_uint, store_int, store_slice, store_grams, store_coins, store_dict, store_maybe_ref}, 
  keywordstyle=[4]\color{brown}\bfseries,
  identifierstyle=\color{black},
  sensitive=false,
  comment=[l]{;;},
  commentstyle=\color{gray}\ttfamily,
  morestring=[b]"
}
\usepackage{pifont}
\usepackage{diagbox}
\usepackage{subfigure}

\usepackage{color}
\usepackage{microtype}

\usepackage{siunitx}
\usepackage{array,framed}

 \usepackage{
   float,
   epsfig,
   wrapfig,
   graphics,
   graphicx,
 }

\usepackage{setspace}
\usepackage{latexsym,url}
\usepackage{enumerate}
\usepackage{algorithm2e}
\usepackage{algpseudocode}
\usepackage{xparse} 
\usepackage{xspace}
\usepackage{multirow}
\usepackage{microtype}

\usepackage{nicematrix}

\usepackage{booktabs}
\usepackage{multirow}
\usepackage{makecell}
\usepackage{enumitem}

\newcommand{\totalDefects}{14,995\xspace}
\newcommand{\overallPrec}{97.49\%\xspace}
\newcommand{\funCDefects}{7,744\xspace}
\newcommand{\funCPrec}{95.09\%\xspace}
\newcommand{\tactDefects}{7,251\xspace}
\newcommand{\tactPrec}{100\%\xspace}

\newcommand{\funCIR}{\textit{FunC IR}\xspace}
\newcommand{\SSAIR}{\textit{SSA-form IR}\xspace}

\newcommand{\defectBadRandomness}{\textit{BR}\xspace}
\newcommand{\defectPrecisionLoss}{\textit{PL}\xspace}
\newcommand{\defectUncheckedReturn}{\textit{UR}\xspace}
\newcommand{\defectGlobalVarRedefined}{\textit{GVR}\xspace}
\newcommand{\defectImproperFunctionModifier}{\textit{IFM}\xspace}
\newcommand{\defectUncheckedBouncedMessage}{\textit{UBM}\xspace}
\newcommand{\defectInconsistentData}{\textit{ID}\xspace}
\newcommand{\defectLackEndParse}{\textit{LEP}\xspace}

\newcommand{\cssize}{\fontsize{5.5pt}{5.6pt}\selectfont}

\usepackage{soul}
\soulregister\cite7 
\soulregister\ref7 

\usepackage{fancyhdr}

\usepackage{subfigure}
\usepackage{lipsum}

\usepackage{xurl} 
\usepackage{tabularx}

\begin{document}

\title{Enhancing The Open Network: Definition and Automated Detection of Smart Contract Defects\\
}

\author{
\IEEEauthorblockN{Hao Song}
\IEEEauthorblockA{\textit{Sichuan University} \\
Chengdu, China \\
ttdelbert@foxmail.com}
\and
\IEEEauthorblockN{Teng Li}
\IEEEauthorblockA{\textit{University of Electronic Science and Technology of China} \\
Chengdu, China \\
tengli2866@gmail.com}
\and
\IEEEauthorblockN{Jiachi Chen*\thanks{*~Jiachi Chen and Ting Chen are the corresponding authors.}}
\IEEEauthorblockA{\textit{Sun Yat-Sen University} \\
Guangzhou, China \\
chenjch86@mail.sysu.edu.cn}
\and
\hspace{1.5cm}
\IEEEauthorblockN{Ting Chen*}
\IEEEauthorblockA{\hspace{1.5cm}\textit{University of Electronic Science and Technology of China} \\
\hspace{1.5cm}Chengdu, China \\
\hspace{1.5cm}brokendragon@uestc.edu.cn}
\and
\hspace{-10cm}
\IEEEauthorblockN{Beibei Li}
\IEEEauthorblockA{\hspace{-10cm}\textit{Sichuan University} \\
\hspace{-10cm}Chengdu, China \\
\hspace{-10cm}libeibei@scu.edu.cn}
\and
\IEEEauthorblockN{Zhangyan Lin}
\IEEEauthorblockA{\textit{University of Electronic Science and Technology of China} \\
Chengdu, China \\
2522579678@qq.com}
\and
\IEEEauthorblockN{Yi Lu}
\IEEEauthorblockA{\textit{BitsLab} \\
Singapore \\
y@movebit.xyz}
\and
\IEEEauthorblockN{Pan Li}
\IEEEauthorblockA{\textit{TonBit} \\
China \\
paul@movebit.xyz}
\and
\IEEEauthorblockN{Xihan Zhou}
\IEEEauthorblockA{\textit{TonBit} \\
China \\
han@movebit.xyz}
}

\maketitle

\begin{abstract}
The Open Network (TON), designed to support Telegram's extensive user base of hundreds of millions, has garnered considerable attention since its launch in 2022. 
\textit{FunC} is the most popular programming language for writing smart contracts on TON. It is distinguished by a unique syntax compared to other smart contract languages. Despite growing interest, research on the practical defects of TON smart contracts is still in its early stages. 
In this paper, we summarize eight smart contract defects identified from TON's official blogs and audit reports, each with detailed definitions and code examples. Furthermore, we propose a static analysis framework called TONScanner to facilitate the detection of these defects. 
Specifically, TONScanner reuses \textit{FunC} compiler's frontend code to transform the \textit{FunC} source code into \textit{FunC} intermediate representation (IR) in the form of a directed acyclic graph (DAG).
Based on this IR, TONScanner constructs a control flow graph (CFG), then transforms it into a static single assignment (SSA) form to simplify further analysis. TONScanner also integrates Data Dependency, Call Graph, Taint Analysis, and Cell Construct, which are specifically tailored for TON blockchain's unique data structures. These components finally facilitate the identification of the eight defects.
We evaluate the effectiveness of TONScanner by applying it to 1,640 smart contracts and find a total of \totalDefects defects. Through random sampling and manual labeling, we find that TONScanner achieves an overall precision of \overallPrec. The results reveal that current TON contracts contain numerous defects, indicating that developers are prone to making errors. TONScanner has proven its ability to accurately identify these defects, thereby aiding in their correction.

\end{abstract}


\section{Introduction}
The Open Network (TON) blockchain \cite{ton} is designed to make blockchain technology~\cite{blockchain_tech} a readily accessible tool for global users~\cite{uniqueaspects}. It aims to be capable of handling millions of transactions per second \cite{tonwhitepaper}, thereby becoming a ubiquitous decentralized application (DApp) platform. Leveraging the extensive user base of Telegram, TON blockchain has garnered significant attention since its testnet phase. Following the launch of the mainnet in 2022, TON blockchain has attracted a substantial number of users interacting with DApps on the platform. As of July 2024, according to DefiLlama, the Total Value Locked (TVL) on TON blockchain has reached \$752.95 million \cite{totalvaluelocked}. Compared to the beginning of the year, TVL on TON blockchain has increased approximately 55 times.

The backend of DApps on TON blockchain is developed based on smart contract technology. Smart contracts are Turing-complete programs that run on the blockchain, allowing users to interact with the decentralized ledger by invoking these contracts \cite{turingcomplete,meta,voting_sys}. 
TON blockchain uses ``sharding'' to horizontally scale overall system performance. This design differentiates TON from traditional blockchain platforms, as TON smart contracts do not perform atomic, synchronous access to persistent states~\cite{sharding}. Instead, the communication paradigm between smart contracts has shifted to a non-atomic, asynchronous approach. To meet performance and scalability requirements, TON innovatively introduces the cell as its basic data structure \cite{cell_overview}. The storage and retrieval of smart contract data in TON are centered around the cell. TON smart contracts also break the limitation of immutable code by providing a native upgrade mechanism \cite{upgrade_api}.
Due to the non-atomicity of calls, unique data structure, and modifiability of TON blockchain, there are significant differences in the implementation of smart contracts compared to traditional blockchain platforms like Ethereum.
Given the unique characteristics of the TON blockchain, developers may face an increased risk of contract defects during the development process. Contract defects refer to errors, flaws, or bugs within smart contracts that result in incorrect or unexpected outcomes, or cause the contract to behave in unforeseen ways \cite{defectdef}. Despite numerous previous studies reporting various smart contract defects and some works proposing frameworks or tools for defect detection, most of this research has focused primarily on Ethereum \cite{mutationtest,DefectChecker,defectdef,ReentrancyDetection,he2024large}. To the best of our knowledge, there is currently no research specifically addressing smart contract defects on TON blockchain. \par

To address this research gap, we first conduct an empirical study to define smart contract defects. 
Specifically, we collect blogs from the official
documentations~\cite{ton_docs}, the official blogs~\cite{ton_blog} and the research community~\cite{ton_community} recommended by the official sources. Additionally, we gather audit reports from blockchain security companies. These sources provide substantial guidance for our defect definitions. Based on open card sorting approach \cite{opencardsorting}, we ultimately define eight types of \textit{FunC} \cite{func} smart contract defects (\textit{FunC} is the smart contract language of TON blockchain, which will be introduced in Section \ref{sec:background}): \textit{Bad Randomness}, \textit{Precision Loss}, \textit{Unchecked Return}, \textit{Global Var Redefined}, \textit{Improper Function Modifier}, \textit{Unchecked Bounced Message}, \textit{Inconsistent Data}, and \textit{Lack End Parse}. 
The first three defects may also appear on other blockchains, while the latter five are specific to TON.\par

Furthermore, to identify the eight defined TON smart contract defects, we propose a static analysis framework called TONScanner. Specifically, TONScanner leverages the \textit{FunC} compiler to transform smart contract source code into a \textit{FunC} intermediate representation (\funCIR) with a directed acyclic graph (DAG). Using \funCIR, TONScanner constructs a Control Flow Graph (CFG)~\cite{cfg}. To enhance analysis efficiency, TONScanner transforms the non-Static Single Assignment (SSA) form CFG into SSA form, resulting in \SSAIR. TONScanner integrates the Data Dependency, Call Graph, Taint Analysis~\cite{taint}, and a customized Cell Construct as analyzers. Utilizing these four analyzers, TONScanner builds eight detectors for the eight previously defined defects. 

Based on the requirements for open-source and verified code, we collect 922 real-world \textit{FunC} contracts and 718 \textit{Tact} contracts (\textit{Tact} \cite{tact} is built on \textit{FunC}, details see Section \ref{sec:background}), totaling 1,640 contracts. Running our framework on these two datasets, we find a total of \totalDefects defects, indicating that the defined defects are prevalent on TON blockchain. To evaluate the performance of TONScanner, we randomly sample the two datasets, using a confidence interval of 10 and a confidence level of 95\%. We then manually label them and compare the results with our detection outcomes. TONScanner achieves precision rates of \funCPrec and \tactPrec on the \textit{FunC} and \textit{Tact} datasets, respectively, with an overall precision of \overallPrec.

The main contributions of our work are as follows:\par
\begin{itemize}
\item We define eight types of \textit{FunC} smart contract defects, including five of them are specific to TON blockchain. For each defect, we provide a code example to help developers understand and avoid these defects.

\item We develop a framework named TONScanner, based on static analysis, for detecting smart contract defects on TON blockchain. This framework transforms source code into a \SSAIR, enabling efficient and accurate defect detection. TONScanner is also highly scalable, supporting the addition of new defect detectors to address future updates and newly identified defects.

\item We test 1,640 smart contracts with TONScanner, finding a total of \totalDefects defects. Additionally, our framework achieves an overall precision of \overallPrec on the manually labeled datasets that we randomly sample from these smart contracts. We publish datasets, raw data, and TONScanner at \href{https://anonymous.4open.science/r/TON_support_material-ECD4/}{\color{NavyBlue}{https://anonymous.4open.science/r/TON\_support\_material-ECD4/}}.

\end{itemize}

\section{Background}
\label{sec:background}

\subsection{TON Smart Contract and TVM}
TON smart contracts are computerized transaction protocols running on TON blockchain~\cite{szabo1997formalizing}, designed to manage distributed ledgers~\cite{ledger} and system state. The system state changes of TON blockchain are governed by a specific execution model \cite{executionmodel}. Similar to Ethereum~\cite{he2024nurgle,he2023tokenaware}, the execution model is implemented through a virtual state machine, known in TON blockchain as TON Virtual Machine (TVM) \cite{evm,tvm}. TVM functions as a quasi-Turing machine, with the total computation limited by gas \cite{gas}. TVM executes smart contracts by representing contract code and data as cells \cite{cell} stored on the blockchain. During execution, TVM loads these cells into the stack and uses control registers to manage the execution state. As a stack machine, TVM performs instructions by manipulating values at the top of the stack, modifying persistent data, and generating output actions \cite{tvmcontract}. TON smart contracts can be written in three languages: \textit{Tact}~\cite{tact}, \textit{FunC}~\cite{func}, and \textit{Fift}~\cite{fift}. Next, we provide explanations of the relevant concepts in this work. \par 

\noindent{\bf FunC.} \textit{FunC} is a high-level, C-like, statically typed language. \textit{FunC} programs are compiled into \textit{Fift} assembly code, which is then converted into bytecode for execution by the TVM.

\noindent{\bf Tact.} 
\textit{Tact} is the second most popular language for TON contract development, after \textit{FunC}. \textit{Tact} is built on \textit{FunC}, and its compiler translates \textit{Tact} programs into \textit{FunC} form. Therefore, our defect definitions are based on \textit{FunC}.

\noindent{\bf recv\_internal Function.} In TON blockchain, this function is executed when a contract is accessed directly~\cite{solidity_vs_func}. 
It can be simply understood as the main function and is a common practice in TON contracts.

\noindent{\bf Bounced Message.} Bounced message is a mechanism where a message that could not be successfully processed is returned to the sender. This is primarily used for error notifications and refunding funds.

\subsection{TON vs. Ethereum}
In this subsection, we compare the differences between TON and Ethereum to facilitate the understanding of this work. We do not delve into the consensus mechanisms or architectures; instead, we focus solely on the differences related to smart contracts, aligning with the theme of our research.\par

\noindent{\bf Non-atomicity of Calls.} On Ethereum, calls between contracts are atomic. If any step in the call chain fails, the entire transaction will be rolled back. However, on TON blockchain, calls between smart contracts are non-atomic \cite{uniqueaspects}. This means that if the final call in a call chain encounters an error, the changes made by previous calls will not be rolled back to their initial state. Thus, developers must account for such scenarios when writing smart contracts. Specifically, developers need to handle messages that are bounced back due to call errors. If a developer fails to handle these bounced messages, it triggers what we define as the \textit{Unchecked Bounced Message} defect.\par

\noindent{\bf Unique Data Structure.} TON blockchain uses a unique data structure called a ``cell'' to store data. A cell can contain up to 1023 bits of data and four references to other cells. Cells are designed as opaque objects optimized for compact storage, making direct modification or reading within smart contracts impossible. Consequently, developers must use a \textit{builder} to add data and create a cell, and a \textit{slice} to read and parse the data from a cell within the smart contract \cite{cell_overview}. This pattern is frequently seen in the \textit{FunC} language used for writing TON smart contracts, and examples of this usage can be found in the code examples in Section \ref{sec:defects}.

\noindent{\bf Modifiability.} Once a smart contract is deployed on Ethereum, it cannot be modified~\cite{yang2023definition}. This immutability forces developers to adopt complex design patterns to upgrade smart contracts, such as the proxy contract pattern~\cite{proxycontract}. This pattern allows developers to point a proxy contract to the address of a new implementation contract, effectively ``upgrading'' the contract without changing its address or state. In contrast, on TON blockchain, smart contracts can be modified. TON standard library defines update functions \cite{standardlibrary}. This feature allows developers to quickly modify smart contracts upon detecting relevant defects when using our tool, thereby preventing irreversible losses.

\section{Defects in TON Smart Contracts}
\label{sec:defects}
\subsection{Data Collection}

\subsubsection{Official Blogs}
The blogs are sourced from official documentations~\cite{ton_docs}, official blogs~\cite{ton_blog}, and TON research community~\cite{ton_community}. These blogs are usually written by security experts or TON contract developers; thus, they typically summarize common issues and provide security development recommendations. We use the keywords ``func secure'' and ``func develop'' for crawling and eventually collect 82 relevant blogs. These 82 blogs totally contain 1327 security tips~\footnote{The example of the blog and tips can be found at: \href{https://blog.ton.org/secure-smart-contract-programming-in-func}{\color{NavyBlue}{https://blog.ton.org/secure-smart-contract-programming-in-func}} } that are used for further manual analysis.

\subsubsection{Audit Reports}
Security audit reports are conducted by specialized firms and are commissioned by blockchain projects at a significant cost. These reports focus on real-world business needs and highlight issues that could lead to potential losses. 
Thus, we also collect audit reports to identify and define defects. Specifically, we refer to the audit companies listed by the official TON sources~\cite{sap}. After a thorough review, we manually collect all seven relevant reports from Certik \cite{certik} and Quantstamp \cite{quantstamp} for further analysis.

\subsection{Data Analysis}
We collaborate with a blockchain security company that specializes in smart contract audits and invite experts to join our data analysis effort. A total of four individuals participate in this work: two of the paper's authors, each with two years of experience in smart contract development, and two professionals specializing in auditing smart contracts. 

After security experts manually exclude data unrelated to \textit{FunC} smart contract defects, we use an open card sorting approach~\cite{opencardsorting} to analyze and categorize the official blogs and audit reports. Specifically, we create a card for each blog's tip and report, dividing the content into several sections: title, description, and recommendations (defect types). Two authors collaborate on the analysis and classification, which take place in two rounds. In the first round, we randomly select 40\% of the cards. We first read the title and description of each card to understand the associated defects. Then, we review the recommendations to understand how to address the identified defects. Cards without root causes are ignored, and possible defects are categorized. The classification process is supervised by security experts to ensure accuracy.\par

In the second round, we independently analyze and categorized the remaining 60\% of the cards following the same steps as in the first round. Then, we compare the results and discuss the differences with security experts. These discussions lead to either unification or elimination of discrepancies. Finally, we categorize the defects into eight types.\par

\begin{figure}[t]
\setlength{\abovecaptionskip}{-0.5cm}
    \centering
    \includegraphics[width=1\linewidth]{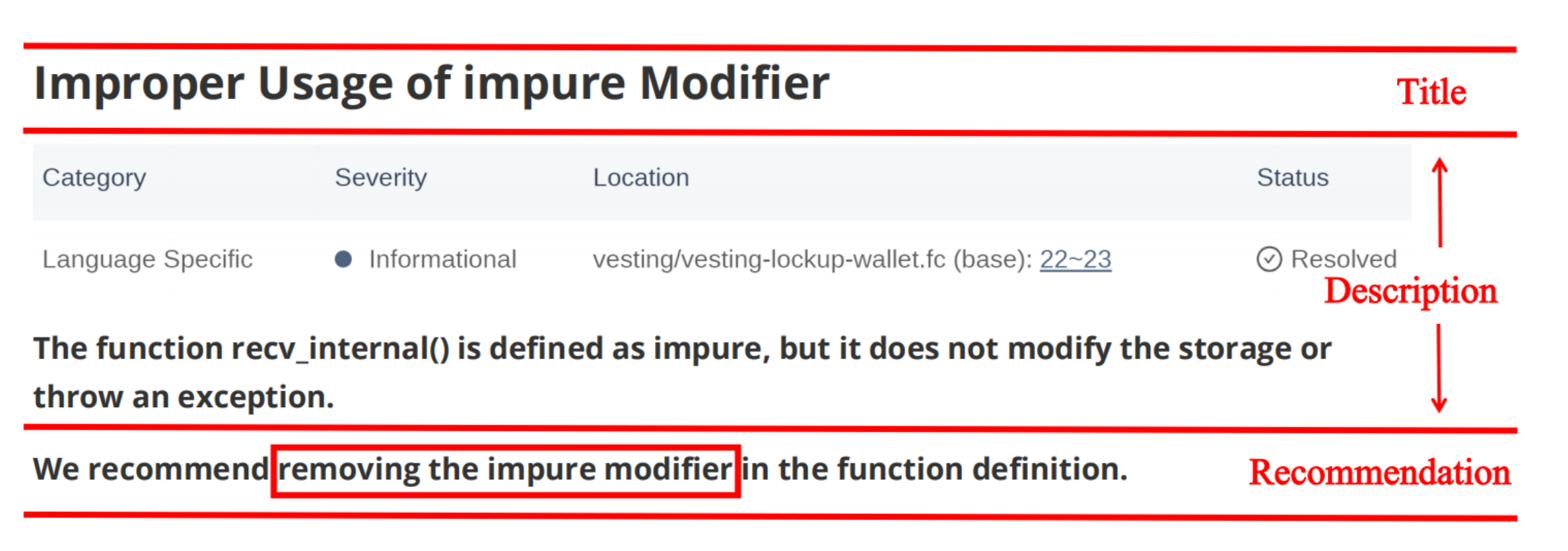}
    \caption{Example of a card of audit reports}
    \label{fig:report}
    \vspace{-0.5cm}
\end{figure}

Figure~\ref{fig:report} illustrates an example of a card generated from a smart contract audit report~\cite{ton_vesting}. The card comprises three parts: the finding name (title), description, and recommendation. As noted in the description, in the \textit{vesting-lockup-wallet.fc} smart contract, the function \textit{recv\_function()} neither modifies the storage nor throws an exception, yet it is annotated with the \textit{impure} modifier. The audit platform recommends removing the impure modifier. Therefore, we categorize this issue as \textit{Improper Function Modifier}.\par

\begin{figure}[t]
\setlength{\abovecaptionskip}{-0.5cm}
    \centering
    \includegraphics[width=1\linewidth]{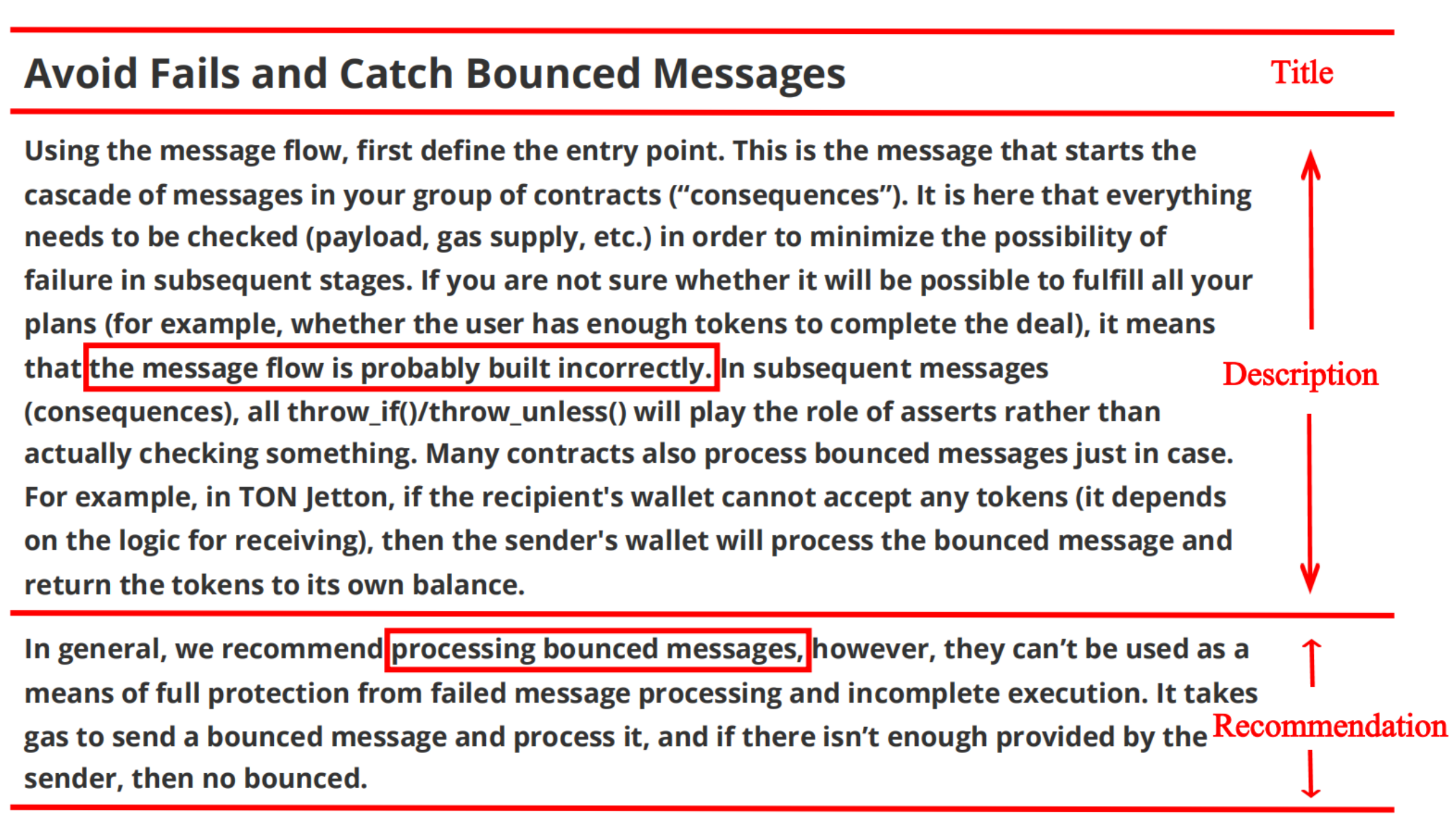}
    \caption{Example of a card of official blogs}
    \label{fig:blog}
    \vspace{-0.5cm}
\end{figure}

Figure~\ref{fig:blog} illustrates a card example generated from a TON official blog tip~\cite{ten_tips}. The card comprises three parts: the tip name (title), description, and recommendation. In the description, it is mentioned that the message flow may be constructed incorrectly, requiring the contract to handle bounced messages. For example, if the receiver's wallet contract is unable to accept tokens, the sender's wallet contract must handle the bounced message. The recommendation similarly suggests processing bounced messages. Therefore, we categorize this issue as \textit{Unchecked Bounced Message}.

\subsection{Defects Definition}
We finally define eight TON smart contracts defects; among them, \textit{Bad Randomness}, \textit{Precision Loss}, and \textit{Unchecked Return} have similar definitions in previous studies on \textit{Solidity} \cite{defectdef,DefectChecker}. Note that there may be other defects common to both \textit{FunC} and \textit{Solidity}. In this paper, we report only those defects that are summarized from official blogs and audit reports to ensure their reliability. Meanwhile, the other five defects are specific to TON. We present a brief definition of each defect in Table \ref{tab:brief_def}, followed by detailed definitions and code examples.

\textbf{(1) Bad Randomness (BR):} Random number generation is commonly used in blockchain projects \cite{randomness}. Similar to traditional programs, generating random numbers in TON blockchain smart contracts also requires a seed. Since the algorithm produces pseudo-random numbers, running the same program multiple times with the same seed will yield the same random number. Developers often use the current logical time as a seed to ensure the randomness. However, similar to Ethereum, miners can influence random number generation, as they determine the current block's information, including the logical time \cite{defectdef}. This makes the generated random numbers still somewhat predictable and controllable.\par

\vspace{-0.1cm}
\begin{lstlisting}[language=FunC, caption=An example of Bad Randomness defect, label=fig:Bad Randomness, firstnumber=1][h]
 () recv_internal (msg_value, in_msg_full, in_msg_body) {
    int seed = cur_lt();
    set_seed(seed);
    if(rand(10000) == 7777) {
        ;;...send reward... 
    }}
\end{lstlisting}
\vspace{-0.1cm}

\textbf{Example:} Listing \ref{fig:Bad Randomness} shows the \textit{recv\_internal} function, designed to implement a  lottery application. 
Among the function parameters, \textit{in\_msg\_full} includes the sender's address. The function body first retrieves the current logical time to use as a random seed (lines 2-3). If the generated random number equals 7777 (line 4), the sender address is rewarded. Since miners can influence the logical time, the random number generation is not entirely unpredictable. This lottery application is unfair, as miners can exploit this defect for profit.\par

\textbf{(2) Precision Loss (PL):} Currently, all unary and binary operators in the \textit{FunC} language are integer operators \cite{precisionloss}. When the result of a division is not an integer, it is rounded down. Thus, executing operations in the order of division followed by multiplication might lead to a deviation from the correct outcome. Although such precision loss may seem insignificant, the improper operation order in decentralized finance (DeFi) \cite{defi} projects can lead to asset loss.\par

\textbf{Example:} In Listing \ref{fig:Precision Loss}, the function \textit{accumulate\_price} is designed to accumulate the calculation of two price metrics based on the passage of time and the values of \textit{ton\_reserve} and \textit{jetton\_reserve} with each call. The value of \textit{time\_elapsed} is obtained by subtracting the timestamp of the last blockchain update from the current time. If \textit{time\_elapsed} is greater than 0 and both {ton\_reserve} and \textit{jetton\_reserve} are non-zero, the code containing the \textit{Precision Loss} defect is executed (lines 5-6). The defective code performs division before multiplying by \textit{time\_elapsed}, leading to coin loss. 
A better approach would be to multiply by \textit{time\_elapsed} first and then perform the division.\par

\vspace{-0.1cm}
\begin{lstlisting}[language=FunC, caption=An example of Precision Loss defect, label=fig:Precision Loss, firstnumber=1][h]
 () accumulate_price() inline {
    int time_elapsed = now() - block_timestamp_last;
    if ((time_elapsed > 0) & (ton_reserve != 0) & 
    (jetton_reserve != 0)) {
        price0_cumulative_last += (ton_reserve / jetton_reserve * time_elapsed);
        price1_cumulative_last += (jetton_reserve / ton_reserve * time_elapsed);
    }
    block_timestamp_last = now();}
\end{lstlisting}
\vspace{-0.1cm}

\begin{table}[t]
\setlength{\abovecaptionskip}{0.1cm}
    \centering
\caption{Definitions of the 8 Defects}
\label{tab:brief_def}
    \begin{tabular}{m{3cm} m{5cm}}
        \toprule
         \textbf{Contract Defect}& \textbf{Definition}\\ 
         \midrule
         \textit{Bad Randomness}& Using transaction logic time as a seed for randomness.\\
         \textit{Precision Loss}& Using an improper order of operations causes result deviation.\\
         \textit{Unchecked Return}& Do not check the return value of call functions.\\
         \textit{Global Var Redefined}& Defining a variable with the same name as a global variable.\\
         \textit{Improper Function Modifier}& Using an incorrect modifier, which causes function call failures.\\
         \textit{Unchecked Bounced Message}& Do not check and handle the bounced message.\\
         \textit{Inconsistent Data}& Reading the data in a manner inconsistent with how it was written.\\
         \textit{Lack End Parse}& Lack of \textit{end\_parse} function to check inputs.\\
         \bottomrule
    \end{tabular}
    \vspace{-0.5cm}
\end{table}

\textbf{(3) Unchecked Return (UR):} This defect refers to not checking return values in the \textit{FunC} smart contracts. The omission might prevent the relevant handling functions from detecting unexpected states and errors. Since calls between smart contracts on TON blockchain are not atomic, a failed call chain does not result in a complete rollback of the state. Developers must use return values to identify and handle exceptions; otherwise, the resulting state from failed transactions can be confusing and misleading.\par

\vspace{-0.1cm}
\begin{lstlisting}[language=FunC, caption=An example of Unchecked Return defect, label=fig:Unchecked Return, firstnumber=1][h]
 () recv_internal(int msg_value, cell in_msg_full, slice in_msg_body) impure {
    slice cs = in_msg_full.begin_parse();
    int flags = cs~load_uint(4);
    ;;if (flags & 1) unchecked flags
    ;;...more code logic...
 }
\end{lstlisting}
\vspace{-0.1cm}

\textbf{Example:} Listing \ref{fig:Unchecked Return} presents a function containing the \textit{Unchecked Return} defect. This function takes a parameter of type \textit{cell} named \textit{in\_msg\_full}. Within the function body, this parameter is converted into a \textit{slice} type for parsing, and a 4-bit unsigned integer is read and stored in the \textit{flag} variable. The \textit{flag} variable holds the return message that needs to be checked, but the code omits the necessary check logic. This results in the contract cannot determine the type of the incoming message, causing it to lose further processing.\par

\textbf{(4) Global Var Redefined (GVR):} References can lead to complex state sharing and modification. Currently, the \textit{FunC} language does not support references. Instead, \textit{FunC} uses global variables to achieve similar functionality. 
If an unexpected assignment operation occurs during a function's execution—specifically assigning a value to a variable with the same name as a global variable—the value of the global variable will be altered.\par

\textbf{Example:} In Listing \ref{fig:Global Var Redefined}, the global variable \textit{tokens} records the total amount. The function \textit{withdraw\_jettons} is used to withdraw tokens and takes a parameter of type \textit{slice}, named \textit{s}. When this function is called, it needs to load the child wallet address and the amount to be withdrawn. Subsequent logic handles the other operations required for the withdrawal. However, while loading the amount, the assigned variable has the same name as the global variable (line 5), causing the total number of tokens (line 1) to be unintentionally altered.\par

\vspace{-0.1cm}
\begin{lstlisting}[language=FunC, caption=An example of Global Var Redefined defect, label=fig:Global Var Redefined, firstnumber=1][h]
 global int tokens;
 () withdraw_jettons(slice src, slice s) impure {
    int query_id = s~load_uint(64);
    slice child_wallet = s~load_msg_addr();
    int tokens = s~load_coins();
    ;;...more code logic...
 }
\end{lstlisting}
\vspace{-0.1cm}

\textbf{(5) Improper Function Modifier (IFM):} In the \textit{FunC} language, function declarations can include modifiers, one of which is the \textit{impure} modifier. This modifier indicates that the function may have significant side effects, such as modifying contract storage, sending messages, or throwing exceptions due to invalid data \cite{impure}. If a function is declared without the \textit{impure} modifier and its return value is not used, the \textit{FunC} compiler might optimize away the function call, assuming that it has no necessary side effects. This feature emphasizes the importance of correctly using modifiers to ensure the intended effects of code are preserved during compilation. This corresponds to a programming paradigm in Ethereum called \textit{State Mutability}, generally using \textit{view} and \textit{pure} to modify functions \cite{solidity_language, state_mutability}.\par

\textbf{Example:} In Listing \ref{fig:Improper Function Modifier}, a function named \textit{store\_base\_data} is defined. This function uses \textit{begin\_cell} to construct a new empty builder and fills it with bank balance, loans, and user data. Finally, it converts the builder into a cell using \textit{end\_cell} (lines 2-6). 
The function employs \textit{set\_data} to write the cell to the contract's storage, thereby modifying the contract's state. However, this modification is rendered ineffective due to the absence of the \textit{impure} modifier annotation.

\vspace{-0.1cm}
\begin{lstlisting}[language=FunC, caption=An example of Improper Function Modifier defect, label=fig:Improper Function Modifier, firstnumber=1][h]
 () store_base_data() {
    set_data(begin_cell()
        .store_coins(ctx_bank_balance)
        .store_coins(ctx_bank_borrowed)
        .store_dict(ctx_users)
        .end_cell());
    commit();}
\end{lstlisting}
\vspace{-0.1cm}

\textbf{(6) Unchecked Bounced Message (UBM):} On TON blockchain, each transaction is executed independently of others, and smart contracts communicate via messages. Due to the asynchronous nature of this communication, developers cannot ensure that message flows will complete as intended. This necessitates thorough checking and handling of bounced messages during contract development to prevent potential issues. Failure to include such processes violates secure development principles and is defined as the \textit{Unchecked Bounced Message} defect.\par

\textbf{Example:} In Listing \ref{fig:Unchecked Bounced Message}, the \textit{recv\_internal} function is used to handle received messages. The function first checks if the message body is not empty before proceeding with further processing. The \textit{Unchecked Bounced Message} defect occurs in the subsequent code logic (lines 6-7). Performing \textit{\&} operation on parameter \textit{flag} and 1 is to determine if the message is a bounced message. However, when this operation yields true, there is no handling logic for the bounced message; instead, the function simply returns. The developer actually loses to capture errors that occur in the middle of the call chain.\par

\vspace{-0.1cm}
\begin{lstlisting}[language=FunC, caption=An example of Unchecked Bounced Message defect, label=fig:Unchecked Bounced Message, firstnumber=1][h]
 () recv_internal(int msg_value, cell in_msg_full, slice in_msg_body) impure {
    if (in_msg_body.slice_empty?()) {
        return ();}  
    slice cs = in_msg_full.begin_parse();
    int flags = cs~load_uint(4);
    if (flags & 1) {
        return ();} 
    if (op == op::withdraw_ton_from_minter) {
        ;;send message with op allows to bounce
    }}
\end{lstlisting}
\vspace{-0.1cm}

\textbf{(7) Inconsistent Data (ID):} TON blockchain stores all content in a data structure called a cell. It requires various types of data to be serialized according to specific rules. The builder and slice primitives provide APIs for storing and reading data in/from cells, respectively \cite{funclib}. These primitives require that the order and type of variables must be consistent when reading and writing the contract's global state. Any inconsistency between them results in the \textit{Inconsistent Data} defect.\par

\textbf{Example:} In Listing \ref{fig:Inconsistent Data}, the \textit{save\_data} function creates a new empty builder and stores a 2-bit unsigned integer named \textit{created\_at} (line 3). However, when reading the data, it attempts to read a 32-bit unsigned integer (line 9). This mismatch causes the written data to be read in an invalid format, resulting in incorrect outcomes.\par

\vspace{-0.1cm}
\begin{lstlisting}[language=FunC, caption=An example of Inconsistent Data defect, label=fig:Inconsistent Data, firstnumber=1][h]
 () save_data() impure {
    set_data(begin_cell()
        .store_uint(created_at, 2)
        .store_uint(state, 2)
        ;;...more code logic...
        .end_cell());}
 () load_data() impure {
    slice data = get_data().begin_parse();
    created_at = data~load_uint(32);
    state = data~load_uint(2);
    ;;...more code logic...
 }
\end{lstlisting}
\vspace{-0.1cm}

\textbf{(8) Lack End Parse (LEP):} Including the \textit{end\_parse} function after each slice read is a best practice in secure \textit{FunC} contract development~\cite{best_practice}. 
The \textit{end\_parse} function checks if the slice is empty after loading its contents. If not, the function throws an exception to ensure that the slice contains expected data. Since omitting the \textit{end\_parse} function does not immediately cause obvious security issues, it may be overlooked by developers. However, TON blockchain uses bit streams with variable data format, using \textit{end\_parse} to ensure that the read and write operations are consistent is helpful \cite{secure_programming}. Therefore, the absence of the \textit{end\_parse} check is defined as a defect.\par

\vspace{-0.1cm}
\begin{lstlisting}[language=FunC, caption=An example of Lack End Parse defect, label=fig:Lack End Parse, firstnumber=1][h]
 () load_data() impure {
    var ds = get_data().begin_parse();
    owner = ds~load_msg_addr();
    flag = ds~load_int(32); 
    ;; ds.end_parse();
 }
\end{lstlisting}
\vspace{-0.1cm}

\textbf{Example:} In Listing \ref{fig:Lack End Parse}, after reading the final 32-bit integer from the slice, the \textit{end\_parse} function should be included to verify that all data has been used (line 5). However, the developer omitted this step (as indicated by line 5), which is necessary to ensure that no unused data remains in the slice. Without this check, the program loses the guarantee that it reads as much data as it writes, which may increase the difficulty of debugging.\par

\section{Methodology}
\label{sec:methodology}

\subsection{Overview}
\begin{figure}[t]
\setlength{\abovecaptionskip}{-0.5cm}
    \centering
    \includegraphics[width=1\linewidth]{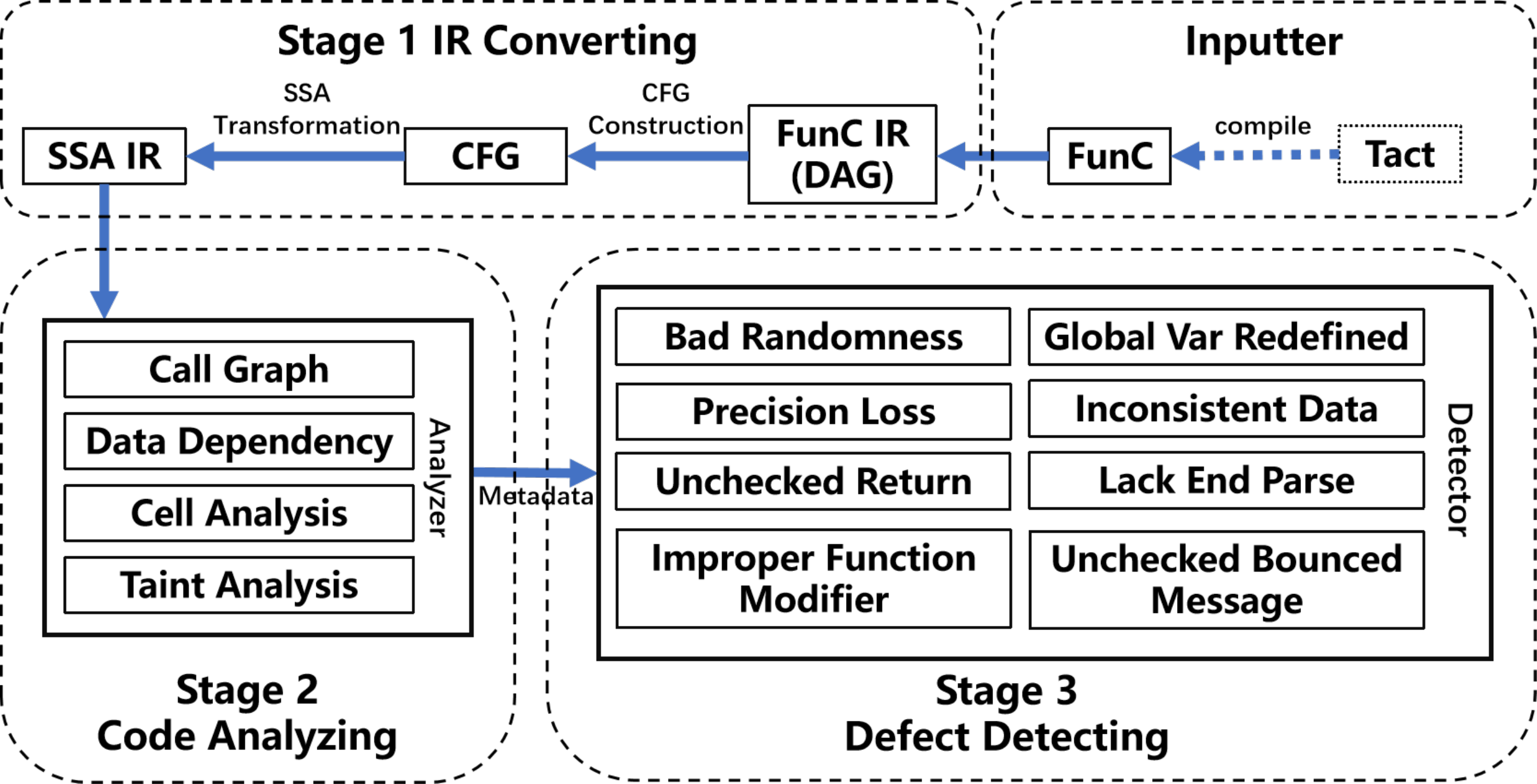}
    \caption{Architecture of TONScanner}
    \label{fig:Architecture}
    \vspace{-0.5cm}
\end{figure}

Figure \ref{fig:Architecture} illustrates the methodology of TONScanner, which performs defect detection through three key stages: \textit{IR Converting}, \textit{Code Analyzing}, and \textit{Defect Detecting}. TONScanner also incorporates an \textit{Inputter} module, which standardizes smart contract source code into \textit{FunC} language format. Specifically, TON smart contracts can be written in \textit{Tact} or \textit{FunC}. For contracts written in \textit{Tact}, TONScanner compiles them into \textit{FunC} code. The differences between these two languages are detailed in Section \ref{sec:background}.\par

To facilitate further detection, TONScanner generates IR in the first stage. \textit{Inputter} module feeds the \textit{FunC} source code into the compiler. The frontend of the \textit{FunC} compiler performs lexical analysis, syntax analysis, and semantic analysis to generate the \funCIR in the form of a DAG. The \textit{FunC} compiler also performs necessary optimizations, such as constant propagation. Next, the DAG-form \funCIR is converted into a CFG. Finally, the \funCIR is transformed into \SSAIR based on the CFG.\par

During the second stage, the \SSAIR serves as input, and the necessary metadata for defect detection is collected. This includes the Call Graph, Data Dependency Tree, Cell Analysis, and Taint Analysis.\par

In the third stage, the metadata collected by the analyzer is used to implement the eight defect detectors. 
These detectors utilize the source map to accurately identify defects and produce a defect detection report.
The architecture of TONScanner is designed to be extensible, allowing detectors to be added without modifying the original design.

\subsection{Stage 1: IR Converting}
In this stage, TONScanner converts the \funCIR into the \SSAIR for further analysis in stage 2.

\subsubsection{CFG}
\begin{figure}[t]
\setlength{\abovecaptionskip}{-0.5cm}
    \centering
    \includegraphics[width=1\linewidth]{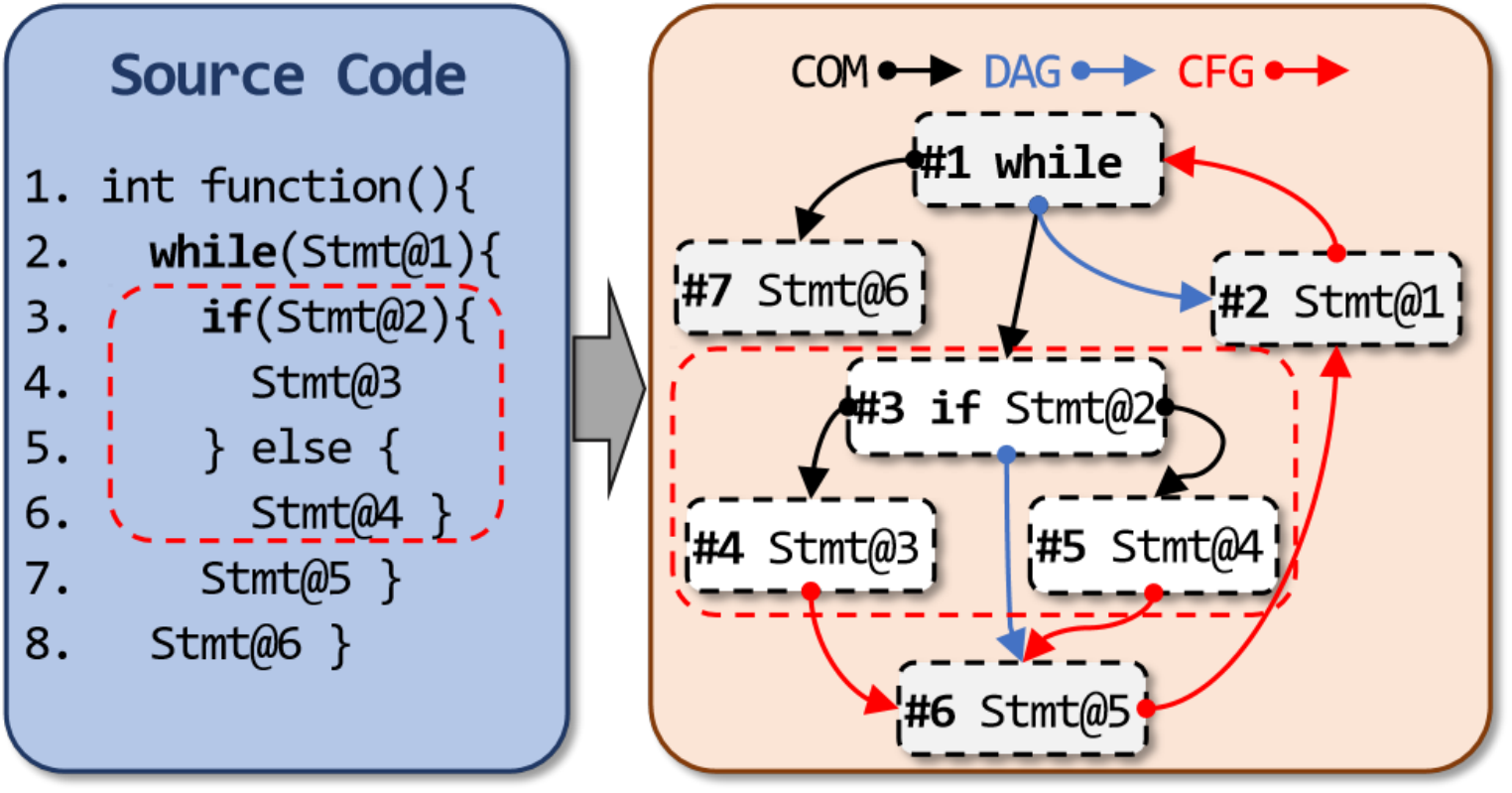}
    \caption{Conversion of DAG to CFG}
    \label{fig:DAG_CFG}
    \vspace{-0.6cm}
\end{figure}

TONScanner converts DAG-form \funCIR to CFG to facilitate further analysis. Figure \ref{fig:DAG_CFG} shows the source code for a function with its DAG and CFG. In this example, blocks \#3, \#4, and \#5 represent the branching structure of lines 3-6 in the source code, while blocks \#1 to \#6 represent the entire loop structure of lines 2-7. Block \#1 and block \#2 form a loop, which is not reflected in the DAG. In contrast, the CFG clearly includes control flow, making it more suitable for analysis.\par

The transition from DAG to CFG focuses on control flow statements, namely conditional branches and loops. For conditional branches, we identify and create basic blocks \cite{bb} within the program: the condition check block, the true condition block (then block), the false condition block (optional else block), and the merge block. The condition check block directs execution to the then block or else block based on the condition's result. The merge block represents code after the branch, linking all blocks to form the CFG. For loop statements, the CFG includes the loop condition check block, loop body block, and exit block. The loop condition check block branches to the loop body block or the exit block, and the loop body block returns to the condition check block, forming a loop back edge. \par

\subsubsection{\SSAIR}
The difference between SSA and non-SSA form lies in variable assignment and management. In SSA form, each variable is assigned exactly once, with new variable versions and $\phi$ functions introduced to handle control flow merge points. In contrast, non-SSA form allows variables to be assigned multiple times, making variable value tracking and management more complex. SSA form simplifies the relationships between variable definitions and uses, making data flow analysis more intuitive and accurate. Additionally, in non-SSA form, data flow analysis requires maintaining the value information for all elements at each program point. If the CFG is large or there are many results to maintain, the space and time overhead can be significant.\par

We use algorithms \cite{ssa} to convert a non-SSA form CFG to SSA form. First, TONScanner analyzes the CFG to determine the dominance frontier for each variable definition. The dominance frontier helps identify control flow convergence points, requiring $\phi$ functions to merge variable values. At these frontiers, $\phi$ functions are inserted to handle merging values from different paths. Next, TONScanner traverses basic blocks to perform reaching-def analysis, which identifies variable definitions reaching each program point, ensuring correct variable versions are used. Finally, all variables are renamed so each is assigned only once in SSA form. This involves updating $\phi$ functions to reflect new variable names, simplifying variable usage tracking throughout the program.

\subsection{Stage 2: Code Analyzing}
In this stage, TONScanner extracts the necessary metadata for further defect detection based on the CFG.

\subsubsection{Call Graph Constructing}

\begin{figure}[t]
\setlength{\abovecaptionskip}{-0.5cm}
    \centering
    \includegraphics[width=1\linewidth]{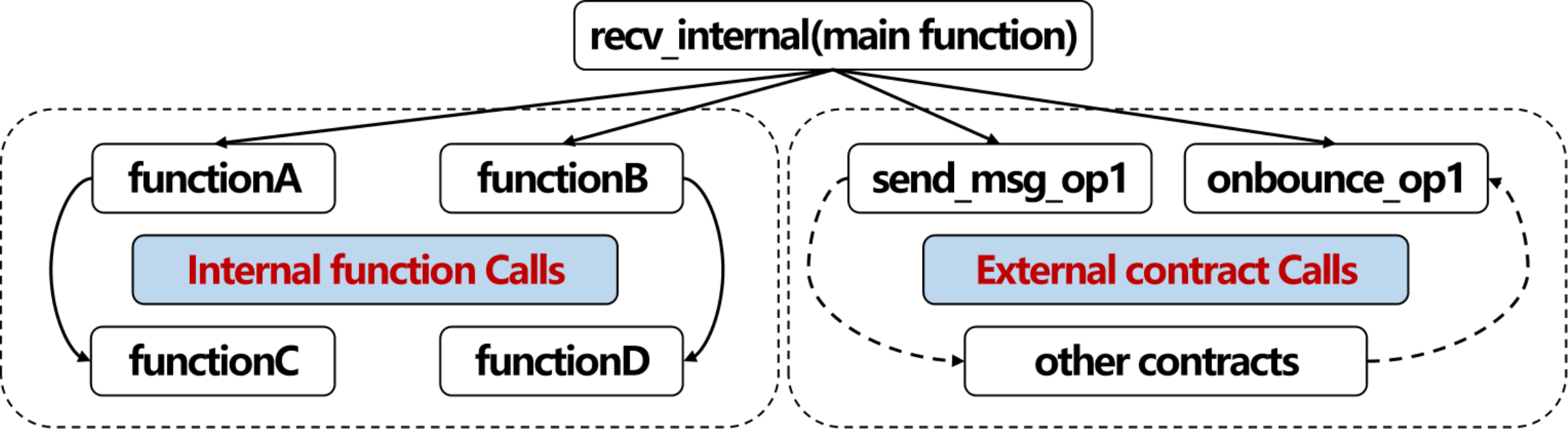}
    \caption{An example of a Call Graph in TON contract}
    \label{fig:Call_Graph}
    \vspace{-0.6cm}
\end{figure}

A general call graph records relationships between internal function calls within a contract. Given the interoperability of smart contracts, inter-contract calls are common. Thus, the call graph here also considers calls to other contracts. Figure \ref{fig:Call_Graph} illustrates the function calls in TON contracts. The left part shows internal function calls, and the right part shows special method calls to external contracts. Due to TON's unique message mechanism, where external contracts are invoked by sending messages, TONScanner includes functions for sending messages and receiving bounced messages as nodes in the call graph.\par
TONScanner constructs the call graph through the following four steps. First, it identifies all functions within the contract to construct the nodes of the call graph. Second, it scans the contract to identify all function call instructions, creating an edge between the calling function and the called function to represent the function call relationship. Third, it scans the contract for instructions that call the \textit{send\_raw\_message} function, which is used to send a message to another contract. This creates a send message node and an edge to represent the message transmission between contracts. Fourth, if the contract handles bounced messages, it can receive a bounced message from an external source, creating a receive bounced message node and an edge to represent the bounced message.

\subsubsection{Data Dependency Tree}
TONScanner determines data dependencies through the following steps. First, TONScanner collects variable assignment statements from the \SSAIR and constructs an index mapping variables to their components. For example, the statement \textit{result\_2 := $\hat{\ }$\_+=\_ (result\_2,m\_0)} in a basic block of CFG maps to the source code \textit{result += m}, where \textit{result\_2} is composed of itself and \textit{m\_0}. Using this index, TONScanner can efficiently retrieve the composition of a given variable. Since TONScanner further converts the \funCIR to  \SSAIR, the composition of variables becomes even clearer. Second, TONScanner traverses the variable assignment statements to construct the data dependency tree.

\subsubsection{Cell Analyzing}
In TON programming, commonly used variable types include \textit{int}, \textit{uint}, \textit{cell}, \textit{slice}, \textit{builder}, \textit{tuple} and etc \cite{tontype}. Among these, \textit{cell} is a data structure unique to the TVM. In TON blockchain, different types of data are ultimately persisted as \textit{cell} types. Both storing and loading data from a \textit{cell} should follow a specific order. Besides, \textit{int} and \textit{uint} types are often used with specific bit widths, and \textit{slice} types may also have bit widths. This requires consideration of bit widths during both storing and loading operations. For example, in Listing \ref{fig:Inconsistent Data}, two \textit{uint} types with a width of 2 bits each are stored sequentially. However, during loading, a \textit{uint} type with a width of 32 bits is extracted. The characteristics of \textit{uint} allow the data types stored in the initial state of the cell to be inferred through sequential analysis of the cell's store or load operations.\par

Cell analysis involves storing/loading data within a cell. To store data, a builder object is created using \textit{begin\_cell}, followed by \textit{store} instructions to store data of specified types and lengths. To load data, \textit{begin\_parse} serializes the cell into a slice, from which data of specified types and lengths are loaded. During the CFG traversal, each block's cell data is recorded, and sub-block computations are performed based on this information. When dealing with branch statements, cell structures from different branches may vary and need merging. If the cell results are the same, they are merged; otherwise, results are recorded separately. The cell records variable type, length, and ID. Since the IR is in SSA form, the ID helps locate the variable definition and corresponding source code position. An \textit{end\_parse} flag is used during analysis. It indicates if a cell parsed with \textit{begin\_parse} has been validated with \textit{end\_parse}, aiding in identifying the \textit{Lack End Parse} defect.

\subsubsection{Taint Analyzing}
Taint analysis establishes the taint propagation relationship graph among variables by traversing Data Dependency Tree, handling three types of instructions: assignment instructions, function call instructions and branch instructions. Specifically, for assignment instructions, arithmetic expressions containing tainted data as the right-hand side of an assignment statement directly assign the value to the left-hand side. The taint mark can be directly propagated from the variables in the right-hand side expression to the left-hand side. For function call instructions, when tainted data is passed as an argument to a function, the function's return value will be marked as tainted. For branch instructions, if a tainted variable is used as the branch condition, all assignment statements within the branch block are tainted.

\subsection{Stage 3: Defect Detecting}
In this stage, each detector in TONScanner retrieves the necessary basic metadata and then identifies whether the target contract contains the defined defects. The scalability of TONScanner allows for the development of new analyzers and detectors within the existing framework.

\subsubsection{Bad Randomness}
TONScanner marks block-related information, such as \textit{cur\_lt} and \textit{block\_lt}, as taint sources since these represent logical times. It uses Taint Analysis to track the propagation of these data throughout the program, focusing particularly on their use in random number generation. During analysis, TONScanner monitors whether tainted data reach taint sinks, such as array index accesses and value comparison operations, to identify potential \textit{BR} defects.

\subsubsection{Precision Loss}
TONScanner traverses the function \SSAIR based on the CFG to identify all multiplication instructions. It then uses the Data Dependency Tree to determine the two operands of these multiplication instructions. If any of these operands result from a division operation, the instruction is flagged for potential precision loss due to the sequence of division followed by multiplication.

\subsubsection{Unchecked Return}
TONScanner uses the call graph to check if the called functions have return values. If the value exists, it then uses the Data Dependency Tree to determine if the value is used by other variables.
Considering that the IR generated from the source code includes the original variable names, if an underscore is used to receive the return value, it indicates that the developer intentionally chose to ignore the return value. In such cases, it is not considered an \textit{UR} defect. In summary, TONScanner only checks whether the defined variable receiving the return value is used.

\subsubsection{Global Var Redefined}
TONScanner traverses all instructions in the \SSAIR based on the CFG to identify \textit{SetGlob} instructions, which are used to modify global variables, and \textit{GlobVar} instructions, which are used to declare global variables. 
Since identifying a \textit{GVR} issue at the IR level alone is challenging, the source map locates these statements in the source code. The source code is then checked for variable declaration keywords like \textit{int}, \textit{slice}, \textit{let} and etc. If such keywords are present, the defect is identified.

\subsubsection{Improper Function Modifier}
TONScanner first identifies functions without the \textit{impure} modifier. It then checks the CFG for four operations: (a) throwing exceptions, e.g., \textit{THROWIFNOT}; (b) sending messages, e.g., \textit{send\_raw\_message}; (c) modifying storage, e.g., \textit{set\_data}; and (d) modifying global variables. Next, it uses the call graph to check if called functions have any of these operations. If a function contains these operations and its return value is unused or absent, it flags an \textit{IFM} defect. Unlike the \textit{UR} detector, this one ensures none of the return values are used.

\subsubsection{Unchecked Bounced Message}
TONScanner follows three steps to detect this defect, i.e., checking sent messages, checking bounced messages, and comparing the results.\par

In the first step, TONScanner traverses the function's \SSAIR based on the CFG to identify the \textit{send\_raw\_message} function. Using cell analysis results, it retrieves the sent message formatted as a cell and extracts the bounceable flag located at the third bit \cite{tonmessage}.
If the message cannot bounce, it is skipped. Otherwise, TONScanner extracts the \textit{op} field from the message to determine the function to invoke in the receiver. If the message contains a body, the \textit{op} field is in the first field of the body; if not, the op field, as a uint32 type, follows some header flags.
In summary, the first step requires TONScanner to extract the \textit{op} field from messages that can bounce.\par

In the second step, TONScanner traverses all instructions in the \textit{recv\_internal} function's \SSAIR based on the CFG, extracting the flag (the fourth bit of message \cite{tonmessage}) that indicates whether the bounced message is handled.
As shown in Listing \ref{fig:Unchecked Bounced Message} (lines 7-9), this involves extracting and checking the flag to determine if the bounced message is processed. If this flag is checked within a basic block, TONScanner calculates all blocks dominated by this basic block and extracts the second field of the message body as the \textit{op} field (The first field of the bounced message is the bounced message flag \textit{0xFFFFFFFF}.).
In summary, the second step requires TONScanner to extract the \textit{op} field from bounced messages.\par
In the third step, TONScanner compares the two \textit{op} lists obtained from the previous two steps. 
If an \textit{op} is present only in the first list, TONScanner reports an \textit{UBM} defect.

\subsubsection{Inconsistent Data}
TONScanner traverses all instructions in the \SSAIR based on the CFG to identify the \textit{load\_data} and \textit{set\_data} functions, which are used to read and write contract storage space. Based on the results of cell analysis, it obtains the structure of the cells being set and loaded. It then matches the types and widths of variables field by field to ensure they correspond one-to-one. If any mismatch is found, it is flagged as an \textit{ID} defect. This detector specifically focuses on identifying \textit{ID} defects when \textit{load\_data} and \textit{set\_data} are used to read and write contract storage space.

\subsubsection{Lack End Parse}
TONScanner first identifies basic blocks with no sub-blocks based on the CFG, indicating these blocks can reach the program's end. It then uses cell analysis results to determine the state of all cells at these blocks. If a cell has not been validated with \textit{end\_parse} and is not passed as a parameter to other functions (i.e., it is not further loaded in other functions), TONScanner flags the cell with a \textit{LEP} defect. Finally, TONScanner reports the defect at the last recorded load operation of the cell. This defect is flagged if a cell in any branch is not validated with \textit{end\_parse}, with the corresponding source code location at the last load operation.

\section{Experiment}
\label{sec:experiment}

\subsection{Experimental Setup}
The experiment is conducted on a server running Ubuntu 18.04.5 LTS and equipped with 12 Intel Xeon Gold 6132 CPUs and 120 GB memory.

\subsubsection{Dataset}
We collect TON contract dataset from two sources: TON Verifier \cite{tonverifier} and GitHub. TON Verifier is an open-source platform maintained by the official TON team, featuring verified and deployed smart contracts on TON blockchain. For the open-source dataset from GitHub, we used GitHub's advanced search function to filter repositories using keywords such as file types ``.func'', ``.fc'', ``.tact'', entry functions ``recv\_internal'' and ``TON'' to obtain the final results. Both of them were collected by April 2024, and then we deduplicated all the collected smart contracts. 
We finally obtained 294 contracts from onchain and 628 contracts from GitHub, resulting in a total of \textbf{922 unique \textit{FunC} contracts}. For contracts originally written in \textit{Tact}, we use the \textit{Tact} compiler to compile them into \textit{FunC} form and then apply the same deduplication method as used for \textit{FunC} contracts. This process results in 258 contracts onchain and 460 contracts from GitHub, totaling \textbf{718 unique \textit{Tact} contracts.} 
We divide the dataset into \textit{FunC} and \textit{Tact} categories to facilitate the evaluation. 

\subsubsection{Evaluation Metrics}
In this paper, we summarize the following research questions (RQs):
\begin{itemize}[leftmargin=*]
\item \textbf{RQ1.} 
What is the prevalence of the eight defects in real-world smart contracts?
\item \textbf{RQ2.} 
What is the effectiveness of TONScanner in identifying defects in \textit{FunC} dataset? 

\item \textbf{RQ3.} 
What is the effectiveness of TONScanner in identifying defects in \textit{Tact} dataset? 
\end{itemize}


\subsection{RQ1: Prevalence of Defects}
To answer RQ1, we run TONScanner on 922 \textit{FunC} contracts and 718 \textit{Tact} contracts. The detailed results are shown in Table \ref{tab:result}, where the column \textbf{\# Defects} shows how many relevant defects appear in the collected datasets (it is possible that a contract contains multiple defects).\par
For \textit{FunC} contracts, \defectUncheckedReturn and \defectLackEndParse are the most common defects. TON use the cell for data storage, which allows only sequential access, leading to a significant amount of \defectUncheckedReturn when preceding data is not used. Most users, when encountering the cell for the first time, are unaware of the necessity to use \textit{end\_parse} to verify the cell, which results in a high incidence of \textit{LEPs}. For \textit{Tact} contracts, \defectUncheckedReturn, \defectLackEndParse and \defectInconsistentData are prevalent and occur frequently, both due to defects caused by the \textit{Tact} compiler. Due to the insufficient knowledge of TON's non-atomic and bounced message characteristics, there is often inadequate handling of bounced messages. This oversight results in a significant prevalence of \defectUncheckedBouncedMessage. Other defects occur relatively infrequently. They are typically associated with users' coding practices (e.g., \defectPrecisionLoss and \defectGlobalVarRedefined), specific business logic (e.g., \defectBadRandomness), or they trigger severe errors that are more readily detected (e.g., \defectImproperFunctionModifier).
Overall, TONScanner detects \funCDefects defects in \textit{FunC} contracts, averaging 8.4 defects per contract. For \textit{Tact} contracts, TONScanner detects \tactDefects defects, averaging 10.1 defects per contract. \par

In the \textit{FunC} dataset, we find 1 smart contract containing 6 types of defects, making it the contract with the most defects. There are 33 smart contracts with 5 types of defects, 190 smart contracts with 4 types of defects, 379 smart contracts with 3 types of defects, and 225 smart contracts with 2 types of defects. In the \textit{Tact} dataset, we find 31 smart contracts containing 5 types of defects, making these the contracts with the most defects. There are 510 smart contracts with 4 types of defects, 136 smart contracts with 3 types of defects, and 40 smart contracts with 2 types of defects. According to the TONScanner report, 1,545 smart contracts contain at least one defect, accounting for about 94\% of the collected dataset. This indicates that many TON smart contract developers lack sufficient experience to meet secure development requirements. 

\subsection{RQ2: Evaluation in \textit{FunC} Dataset}
In this RQ, we evaluate the performance of TONScanner based on the \textit{FunC} dataset. Specifically, we randomly sample contracts from the positive results reported for each defect. To determine the sample size for each defect, we follow a confidence interval-based sampling method \cite{confidence_interval} to generalize the total number of issues discovered for that defect. We set a confidence interval of 10 and a confidence level of 95\%, calculating the required number of samples \cite{calculator}. The calculated sample sizes for the eight defects are 2, 46, 94, 16, 4, 72, 80, and 93, respectively. We then sample the evaluation dataset based on these results and carefully manually label them by two authors. During the labeling process, we separate false positives from true positives to analyze the performance of TONScanner on the \textit{FunC} dataset. \par

Columns 4, 6, 8 and 10 of Table \ref{tab:result} summarize the results of applying TONScanner to our manually labeled \textit{FunC} contract samples. Due to the low number of detected \textit{BR}, \textit{GVR}, and \textit{IFM} defects, we analyze all contracts with these three defects to make the results more reliable. Columns 6 and 8 display the number of samples categorized as true positives (TP) and false positives (FP), respectively. We use precision rate to demonstrate the performance of detecting each defect. The precision rate is calculated $\frac{\#TP}{\#TP + \#FP} \times 100\%$. We also calculate the overall precision to demonstrate the effectiveness of TONScanner on the \textit{FunC} dataset. The formula for this calculation is $\frac{\sum_{i=1}^n p_{c_i} \times |c_i|}{\sum_{i=1}^n |c_i|}$, in which $p_{c_i}$ represents the precision of detecting defect $i$, and $|c_i|$ is the number of defect $i$ in the \textit{FunC} dataset.\par

For \textit{UBM}, \textit{ID}, and \textit{LEP} defects, TONScanner reports them at a precision of 93.06\%, 83.75\%, and 97.85\%, respectively. For the remaining 5 defects, our tool achieves 100\% precision. Furthermore, the overall precision of TONScanner achieves \funCPrec on the \textit{FunC} dataset.

\begin{table}[t]\tiny
\setlength{\abovecaptionskip}{-0.1cm}
    \centering
\caption{Experimental Results on \textit{FunC} and \textit{Tact} Datasets}
\label{tab:result}
    \begin{tabular}{l|c|c|c|c|c|c|c|c|c|c} \hline 
         \multirow{2}{*}{\textbf{}} & \multicolumn{2}{c|}{\scriptsize \textbf{\# Defects}} & \multicolumn{2}{c|}{\scriptsize \textbf{\# Samples}} & \multicolumn{2}{c|}{\scriptsize \textbf{\# TP}} & \multicolumn{2}{c|}{\scriptsize \textbf{\# FP}} & \multicolumn{2}{c}{\scriptsize \textbf{Perc(\%)}} \\ \cline{2-11}
          \textbf{} & \cssize \textit{FunC} & \cssize \textit{Tact} & \cssize \textit{FunC} & \cssize \textit{Tact} & \cssize \textit{FunC} & \cssize \textit{Tact} & \cssize \textit{FunC} & \cssize \textit{Tact} & \cssize \textit{FunC} & \cssize \textit{Tact}  \\ \hline
         \cssize \defectBadRandomness & \cssize 2 & \cssize / & \cssize 2 & \cssize / & \cssize 2 & \cssize / & \cssize 0 & \cssize / & \cssize 100 & \cssize / \\ \hline 
         \cssize \defectPrecisionLoss & \cssize 90 & \cssize 53 & \cssize 46 & \cssize 34 & \cssize 46 & \cssize 34 & \cssize 0 & \cssize 0 & \cssize 100 & \cssize 100 \\ \hline
         \cssize \defectUncheckedReturn & \cssize 3927 & \cssize 4338 & \cssize 94 & \cssize 94 & \cssize 94 & \cssize 94 & \cssize 0 & \cssize 0 & \cssize 100 & \cssize 100 \\ \hline 
         \cssize \defectGlobalVarRedefined & \cssize 16 & \cssize / & \cssize 16 & \cssize / & \cssize 16 & \cssize / & \cssize 0 & \cssize / & \cssize 100 & \cssize / \\ \hline 
         \cssize \defectImproperFunctionModifier & \cssize 4 & \cssize 6 & \cssize 4 & \cssize 6 & \cssize 4 & \cssize 6 & \cssize 0 & \cssize 0 & \cssize 100 & \cssize 100 \\ \hline 
         \cssize \defectUncheckedBouncedMessage & \cssize 288 & \cssize 441 & \cssize 72 & \cssize 79 & \cssize 67 & \cssize 79 & \cssize 5 & \cssize 0 & \cssize 93.06 & \cssize 100 \\ \hline 
         \cssize \defectInconsistentData & \cssize 493 & \cssize 976 & \cssize 80 & \cssize 87 & \cssize 67 & \cssize 87 & \cssize 13 & \cssize 0 & \cssize 83.75 & \cssize 100 \\ \hline
         \cssize \defectLackEndParse & \cssize 2932 & \cssize 1618 & \cssize 93 & \cssize 91 & \cssize 91 & \cssize 91 & \cssize 2 & \cssize 0 & \cssize 97.85 & \cssize 100 \\ \hline 
    \end{tabular}
    \vspace{-0.5cm}
\end{table}

\noindent{\bf False Positives Analysis.} 
(1) When parsing messages, contracts handle the \textit{op} field. Contracts evaluate the value of the variable \textit{op} within branch statements to determine different business logics. For \textit{UBM}, TONScanner does not incorporate conditional constraint analysis, resulting in FPs. Moreover, custom message formats may result in incorrectly parsed \textit{op} fields, which is another cause of such FPs.
(2) For \textit{ID}, we find that in branch structures, different conditions execute different logic, making it impossible to account for the cell structure in all cases, leading to false positives. (3) As for \textit{LEP}, we find that false positives occur due to load operations on cells within complex loops, which makes it challenging to accurately identify the cell structure and the usage of \textit{end\_parse}.

\noindent{\bf False Negatives Analysis.} To find contracts with defects that TONScanner fails to report, we follow the same sampling method used for precision analysis, adopting a confidence interval of 10 and a confidence level of 95\%. We sample 87 contracts from the 922 \textit{FunC} contracts and conduct a manual audit to identify defects, resulting in a total of 830 defects. We compare the manually labeled results with the tool's detection results, finding that 25 defects are missed. Among these, 13 are \textit{LEP} defects, and 12 are \textit{UBM} defects. 
(1) When a cell is used within a loop, the number of iterations cannot be determined. This makes it impossible to pinpoint the position of \textit{end\_parse}, resulting in \textit{LEP} defects being missed. (2) When sending messages, contract developers do not always adhere to the officially standardized message format but instead use custom message formats. This leads to message parsing failures, resulting in missed \textit{UBM} defects.

\subsection{RQ3: Evaluation in \textit{Tact} Dataset}
To answer RQ3, we select the same sampling method. The calculated sample sizes for the six defects are 34, 94, 6, 51, 86, and 91, respectively. Two authors then manually label the samples, analyze the results, and calculate the TP and FP. We use the precision rate to evaluate the performance of TONScanner on the \textit{Tact} dataset.

\noindent{\bf False Positive Analysis.} 
TONScanner reports six detected defects with 100\% precision, indicating that it maintains a high detection precision on the \textit{Tact} dataset as well. TONScanner did not detect any \textit{BR} defects because they are rely on specific business scenarios and are relatively rare. The absence of \textit{GVR} is due to the syntax characteristics of \textit{Tact}, which uses member variables of the \textit{Contract} class to store global data, so there are no custom global variables in the generated \textit{FunC} code. Among the six detected defects, only \textit{PL} and \textit{UBM} are controlled by the developer, while the others are caused by the compiler. Through the sampling analysis of the detection results, all these defects are true positives. The four compiler-caused defects show the same defect pattern, indicating that the \textit{Tact} compiler is currently immature and needs optimization.

\noindent{\bf False Negatives Analysis.} To find missed defects in contracts, we use the same sampling method as in the previous section. We sample 85 \textit{Tact} contracts from the 718 in the \textit{Tact} dataset. After manually auditing and labeling the defects, we identify a total of 740 defects. Comparing the labeled results with TONScanner's detection results, we find one missed \textit{LEP} defect. 
The reason for this missed detection is caused by loops.

\section{Discussion}
\label{sec:discussion}

\subsection{Case Studies of FP and FN Instance}

Listing~\ref{fig:fp} is taken from contract~\footnote{Contract can be found at: \href{https://verifier.ton.org/EQCPXKOygMDIuWI6vnSBt498PwHAwaatg5yLYC7TitxLLQbr}{\color{NavyBlue}{https://verifier.ton.org/EQCPXKOygMDIuWI6v\allowbreak-nSBt498PwHAwaatg5yLYC7TitxLLQbr}} }. The code is to extract the necessary information from the incoming message, then construct and forward the message. However, it is flagged as having a \textit{UBM} defect, which is an FP. To distinguish different messages, different \textit{op} codes are used, and the recipient performs different operations depending on the value of the \textit{op} (line 1). When conducting \textit{UBM} defect detection, it is necessary to identify the \textit{op}. However, in this case, the condition \textit{if(op == op::proxy\_send)} imposes a constraint on the \textit{op} within the if-branch. This limits the flexibility of the implementation, as it can only send messages that match the \textit{op} value. In the case of FPs in \textit{UBM} defect detection, to achieve the desired constraint functionality, it requires symbolic execution techniques. \par

\vspace{-0.1cm}
\begin{lstlisting}[language=FunC, caption=FP instance of \textit{UBM} defect, label=fig:fp, firstnumber=1][h]
 if (op == op::proxy_send) {
    ;;...more code logic...
 }
\end{lstlisting}
\vspace{-0.1cm}

Listing~\ref{fig:fn} is taken from contract~\footnote{Contract can be found at: \href{https://github.com/ton-link/ton-link-contract-v3/tree/main/typescript/source/CustomTokenOracle}{\color{NavyBlue}{https://github.com/ton-link/ton-link-contract-v3/tree/main/typescript/source/CustomTokenOracle}} }. This code performs the parsing of a job task and status checking. The \textit{end\_parse} function is missing, but TONScanner encountered an FN issue. The missed \textit{LEP} defects occur because when a cell is used within a loop (lines 3-4), the number of iterations cannot be determined, leading to inaccurate cell analysis. As a result, it is impossible to identify the position of the \textit{end\_parse}. The missed \textit{LEP} defects occur because when a cell is used within a loop, the number of iterations cannot be determined, leading to inaccurate cell analysis. As a result, it is impossible to identify the position of the \textit{end\_parse}.

\vspace{-0.1cm}
\begin{lstlisting}[language=FunC, caption=FN instance of \textit{LEP} defect, label=fig:fn, firstnumber=1][h]
 int job::double_answer(slice, int count, int) impure {
    ;;...more code logic...
    repeat(count) {
        status = status_slice~load_uint(1);}}
\end{lstlisting}
\vspace{-0.1cm}

\subsection{Threats to Validity}
\noindent {\bf Internal Validity.}
To evaluate TONScanner, we need to conduct random sampling checks on the collected smart contract dataset. Individual misunderstandings of smart contracts can lead to inaccuracies in our calculated precision. To reduce errors, we select the most experienced authors to perform the initial checks. After this, we invite security experts to review the results, ensuring the accuracy of our final calculations.

\noindent {\bf External Validity.}
TON has garnered significant attention, leading developers to deploy an increasing number of new smart contracts. Many new TON ecosystem projects may publish their blogs and audit reports, potentially exposing new defects. Our tool adopts a modular design, enabling the development of analyzers and detectors for new defects without affecting the overall architecture. Additionally, it allows for the restructuring of existing analyzers and detectors to adapt to updates.

\subsection{Possible Solutions}
To assist developers in writing secure TON smart contracts, we provide possible defect solutions at Table~\ref{tab:solution} to help developers avoid the defined defects. 
\vspace{-0.35cm}
\begin{table}[h]
\setlength{\abovecaptionskip}{0.1cm}
    \centering
\caption{Possible Solutions for the 8 Defects}
\label{tab:solution}
    \begin{tabular}{m{3cm} m{5cm}}
        \toprule
         \textbf{Contract Defect}& \textbf{Possible Solution}\\
         \midrule
         \textit{Bad Randomness}& Use external sources of randomness via oracles.\\ 
         \textit{Precision Loss}& Allow multiplication to be executed before division.\\ 
         \textit{Unchecked Return}& Check return values every time.\\ 
         \textit{Global Var Redefined}& Avoid defining variables with the same names as global variables.\\  
         \textit{Improper Function Modifier}& Add the \textit{impure} modifier when a function has side effects; otherwise, remove it.\\ 
         \textit{Unchecked Bounced Message}& Catch and process bounced message.\\ 
         \textit{Inconsistent Data}& Ensure that the order and type of read/write variables are consistent each time.\\ 
         \textit{Lack End Parse}& Include the \textit{end\_parse} function after each slice read.\\
         \bottomrule
    \end{tabular}
    \vspace{-0.5cm}
\end{table}

\section{Related Work}
\label{sec:rw}

\subsection{Smart Contracts Defects}
Atzei et al. summarized 12 security vulnerabilities at three levels—\textit{Solidity}, EVM, and blockchain. 
Although this work represents the first comprehensive overview of attacks on Ethereum smart contracts, it does not provide rigorous definitions of the vulnerabilities from a security perspective \cite{atzei2017survey}. Chen et al. used an open card sorting approach to identify and categorize 20 Ethereum smart contract defects from posts collected on StackExchange \cite{defectdef}. In another study, they developed a tool named DefectChecker, which can detect the smart contract defects they defined in their previous work by analyzing the bytecode of smart contracts \cite{DefectChecker}. The research team recently provided precise definitions for five specific defects in NFT smart contracts. Based on these definitions, they developed a tool called NFTGuard, which uses symbolic execution to detect these defined defects \cite{yang2023definition}. Overall, current research on smart contract defect definitions remains focused on the \textit{Solidity} language and the Ethereum ecosystem. Studies on other smart contract development languages and blockchain ecosystems are limited. 
Our work addresses this gap by identifying defects specific to \textit{FunC} smart contracts.

\subsection{Detection Tools for \textit{Solidity} Smart Contracts}
Tools for detecting security issues in \textit{Solidity} smart contracts are currently abundant. Specifically, Luu et al. proposed a tool named Oyente, which detects smart contract vulnerabilities based on symbolic execution \cite{luu2016making}. Similarly, Nikoli{\'c} et al. developed a tool called MAIAN, also based on symbolic execution, that detects smart contract security issues using defined execution rules, similar to Oyente \cite{nikolic2018finding}. Securify decompiles EVM bytecode and analyzes semantic information to detect smart contract security issues, capable of identifying 9 security problems \cite{tsankov2018securify}. Mythril, an industrial tool developed by ConsenSys, constructs a CFG and uses Z3 \cite{z3} as an SMT solver to detect issues based on predefined rules \cite{mythril}. All these tools are developed using static analysis techniques, consistent with the technical background of our work. Some work is based on dynamic techniques. Specifically, ContractFuzzer \cite{contractfuzzer}, ItyFuzz \cite{ityfuzz}, and sFuzz \cite{sfuzz} utilize fuzzing techniques to detect security issues in \textit{Solidity} smart contracts. 

\section{Conclusion}
\label{sec:conclusion}
In this paper, our research addresses a significant gap in the field of smart contract security on TON blockchain, an area previously underexplored compared to platforms like Ethereum. We analyze the collected  official blogs  and audit reports to define 8 defects in TON smart contracts. Additionally, we provide \textit{FunC} code examples for each defect, which helps developers and researchers gain a deeper understanding of these defects. In this paper, we summarize eight defects for TON smart contracts by analyzing official blogs and audit reports. 
We then present TONScanner, a framework designed to accurately detect these TON defects. The tool leverages static analysis techniques to transform source code into a \SSAIR, facilitating efficient and precise detection of various smart contract defects. TONScanner is built with a modular design, allowing developers to customize analyzers and detectors to identify potential new defects.
Our evaluation shows that TONScanner can effectively identify defects in \textit{FunC} and \textit{Tact} smart contracts, discovering a total of \totalDefects defects across both datasets. TONScanner achieved overall precision rates of \funCPrec and \tactPrec on the manually labeled \textit{FunC} and \textit{Tact} datasets, respectively. The overall precision rate on total datasets is \overallPrec.

\section{Acknowledgment}
\label{sec:ack}
The authors thank Ton Foundation and BitsLab for their invaluable support and the anonymous reviewers for their constructive comments. This work is partially supported by the National Natural Science Foundation of China (Grant No. 62332004) and the Sichuan Provincial Natural Science Foundation for Distinguished Young Scholars (Grant No. 2023NSFSC1963).

\normalem
\bibliography{ref}
\bibliographystyle{IEEEtran}

\end{document}